\documentclass[runningheads]{llncs}
\usepackage[T1]{fontenc}
\usepackage{graphicx}
\usepackage{booktabs}
\usepackage{multirow}
\usepackage{amsmath,amssymb,amsfonts}
\usepackage[colorlinks=true,linkcolor=black,citecolor=black,urlcolor=blue,bookmarks=false,hypertexnames=true]{hyperref} 
\begin{document}
\title{$k$-$t$ CLAIR: Self-Consistency Guided Multi-Prior Learning for Dynamic Parallel MR Image Reconstruction}
\titlerunning{$k$-$t$ CLAIR: Self-Consistency Guided Multi-Prior Learning}
%
\author{Liping Zhang\orcidID{0000-0003-1962-0106} \and
	Weitian Chen\orcidID{0000-0001-7242-9285}}
\authorrunning{L. Zhang and W. Chen}
%
\institute{CUHK Lab of AI in Radiology (CLAIR), Department of Imaging and Interventional Radiology, The Chinese University of Hong Kong, Hong Kong\\
	\email{lpzhang@link.cuhk.edu.hk, wtchen@cuhk.edu.hk}\\
    \url{https://github.com/lpzhang/ktCLAIR}}

\maketitle              
\begin{abstract}
Cardiac magnetic resonance imaging (CMR) has been widely used in clinical practice for the medical diagnosis of cardiac diseases. However, the long acquisition time hinders its development in real-time applications. Here, we propose a novel self-consistency guided multi-prior learning framework named $k$-$t$ CLAIR to exploit spatiotemporal correlations from highly undersampled data for accelerated dynamic parallel MRI reconstruction. The $k$-$t$ CLAIR progressively reconstructs faithful images by leveraging multiple complementary priors learned in the $x$-$t$, $x$-$f$, and $k$-$t$ domains in an iterative fashion, as dynamic MRI exhibits high spatiotemporal redundancy. Additionally, $k$-$t$ CLAIR incorporates calibration information for prior learning, resulting in a more consistent reconstruction.
Experimental results on cardiac cine and T1W/T2W images demonstrate that $k$-$t$ CLAIR achieves high-quality dynamic MR reconstruction in terms of both quantitative and qualitative performance.
\end{abstract}
\section{Introduction}
Cardiac magnetic resonance imaging (CMR) has gained widespread adoption in the diagnosis of cardiac diseases, owing to its exceptional soft-tissue contrast, non-invasive nature, and high spatial resolution. However, the acquisition procedure is inherently time-consuming due to the requirement of repeated acquisition of multiple heartbeat cycles. The prolonged duration can lead to potential patient discomfort and introduce motion artifacts in the resulting images.

Undersampling k-space is effective for accelerating the process, but direct reconstruction using incomplete data results in poor signal-to-noise ratio (SNR) and severe artifacts due to the violation of the Nyquist sampling theorem. Parallel imaging (PI) techniques~\cite{pruessmann1999sense,griswold2002generalized} have been widely adopted to achieve acceleration in MR acquisition. However, their speedup rates are often constrained by hardware limitations.
The combination of compressed sensing (CS) techniques with PI methods~\cite{lustig2007sparse,lustig2010spirit,jung2009k,lingala2011accelerated,otazo2015low} has shown promise in enabling rapid imaging at high acceleration rates and improving the quality of image reconstruction for accelerated dynamic MRI. However, CS-based methods often rely on the assumption that the desired image has a sparse representation and incoherent undersampling artifacts in a known transform domain. 

Deep learning (DL)-based reconstruction approaches have gained significant popularity in recent years, surpassing the limitations of traditional CS techniques. These methods utilize neural networks, either directly or iteratively, to address the inverse problem in either the image domain~\cite{schlemper2017deep,hammernik2018learning,aggarwal2018modl,yang2017dagan} or the k-space domain~\cite{han2020k,akccakaya2019scan}. However, a common issue with these approaches is the excessive smoothing of reconstructed images. To tackle this challenge, recent advancements have focused on cross-domain models~\cite{eo2018kiki,ran2020md,wang2022dimension,fabian2022humus,zhang2023camp}, which leverage multiple sources of prior knowledge to achieve high-quality image reconstruction while preserving sharp details. However, most existing methods in the literature primarily focus on static MRI and do not adequately address the challenges of accelerated dynamic MRI reconstruction, which requires capturing crucial spatial-temporal information. Some recent approaches~\cite{schlemper2017deep,qin2018convolutional,qin2019k} have emerged to exploit temporal correlations between dynamic MR frames in the spatiotemporal domain. However, these methods have mainly focused on single-coil cardiac MR reconstruction, with only a limited number of methods available for multi-coil cardiac MR reconstruction~\cite{kustner2020cinenet,qin2021complementary}. As a result, there is a pressing need for more efficient and effective deep learning models to further advance fast cardiac MR.

In this work, we propose $k$-$t$ CLAIR, a self-consistency guided multi-prior learning framework for accelerated dynamic parallel MRI reconstruction. It leverages unrolled neural networks to exploit spatiotemporal correlations by learning complementary priors in different domains: spatiotemporal ($x$-$y$-$t$ space), frequency-temporal ($k_x$-$k_y$-$t$ space), and spatial-temporal frequency ($x$-$y$-$f$ space). Self-consistency learning is enforced in the $k$-$t$ domain using an end-to-end data-driven approach. Additionally, we introduce a frequency fusion layer to coordinate feature learning across all priors, facilitating faithful dynamic MRI reconstruction. Experimental results on highly undersampled cardiac cine, T1 weighted (T1W) acquisitions for T1 mapping, and T2 weighted (T2W) acquisitions for T2 mapping demonstrate the superior performance of our proposed method in reconstructing high-quality dynamic images across various accelerations.

\section{Methods}
\subsection{Dynamic Parallel MRI Problem Formulation}
In Parallel MRI, a complex-valued MR image sequence denoted as $\mathbf{m} \in \mathbb{C}^N$ in the $x$-$y$-$t$ space is simultaneously encoded by $N_c$ receiver coils. The acquired k-space measurement, represented as $\tilde{\mathbf{v}} \in \mathbb{C}^{N_cN}$ in the $k_x$-$k_y$-$t$ space, can be expressed as:
\begin{equation}
	\tilde{\mathbf{v}} = \mathcal{M} \mathcal{F}_\text{s}\mathcal{S}\mathbf{m},
	\label{eq1}
\end{equation}
where the forward process involves sequential operators of coil sensitivity map projection $\mathcal{S}$, spatial Fourier transform $\mathcal{F}_\text{s}$, and sampling pattern $\mathcal{M}$. The reconstruction of $\mathbf{m}$ from $\tilde{\mathbf{v}}$ is an ill-posed inverse problem. While conventional CS-MRI methods~\cite{lustig2007sparse} utilize predefined prior knowledge to solve the inverse problem, we formulate it as a learnable multi-prior optimization problem:
\begin{equation}
	\mathop{\min}\limits_\mathbf{m}
	{
		\| \mathcal{A}\mathbf{m} - \tilde{\mathbf{v}} \| ^2_2
		+ \lambda_\text{xt} \mathcal{R}_\text{xt}(\mathbf{m};\theta_\text{xt})
		+ \lambda_\text{xf} \mathcal{R}_\text{xf}(\mathcal{F}_\text{t}\mathbf{m};\theta_\text{xf})
		+ \lambda_\text{kt} \mathcal{R}_\text{kt}(\mathcal{F}_\text{s}\mathbf{m};\theta_\text{kt})
	},
	\label{eq2}
\end{equation}
where
$\mathcal{R}_\text{xt}(\mathbf{m};\theta_\text{xt})$, $\mathcal{R}_\text{xf}(\mathcal{F}_\text{t}\mathbf{m};\theta_\text{xf})$, and
$\mathcal{R}_\text{kt}(\mathcal{F}_\text{s}\mathbf{m};\theta_\text{kt})$
are data-adaptive priors with learnable parameters $\theta_\text{xt}$, $\theta_\text{xf}$, and $\theta_\text{kt}$ to regularize the data in the $x$-$t$, $x$-$f$, and $k$-$t$ spaces, respectively. $\mathcal{F}_\text{t}$ and $\mathcal{F}_\text{s}$ represent the Fourier transform operators along the temporal ($t$) and spatial ($x$-$y$) dimensions of the image sequence $\mathbf{m}$, respectively. The parameters $\lambda_\text{xt}$, $\lambda_\text{xf}$, and $\lambda_\text{kt}$ control the balance between the impact of the imposed prior regularization and the fidelity to the acquired data. 

\subsection{$k$-$t$ CLAIR for Dynamic Parallel MRI Reconstruction}
\begin{figure}[!t]
	\centering
	\includegraphics[width=\textwidth]{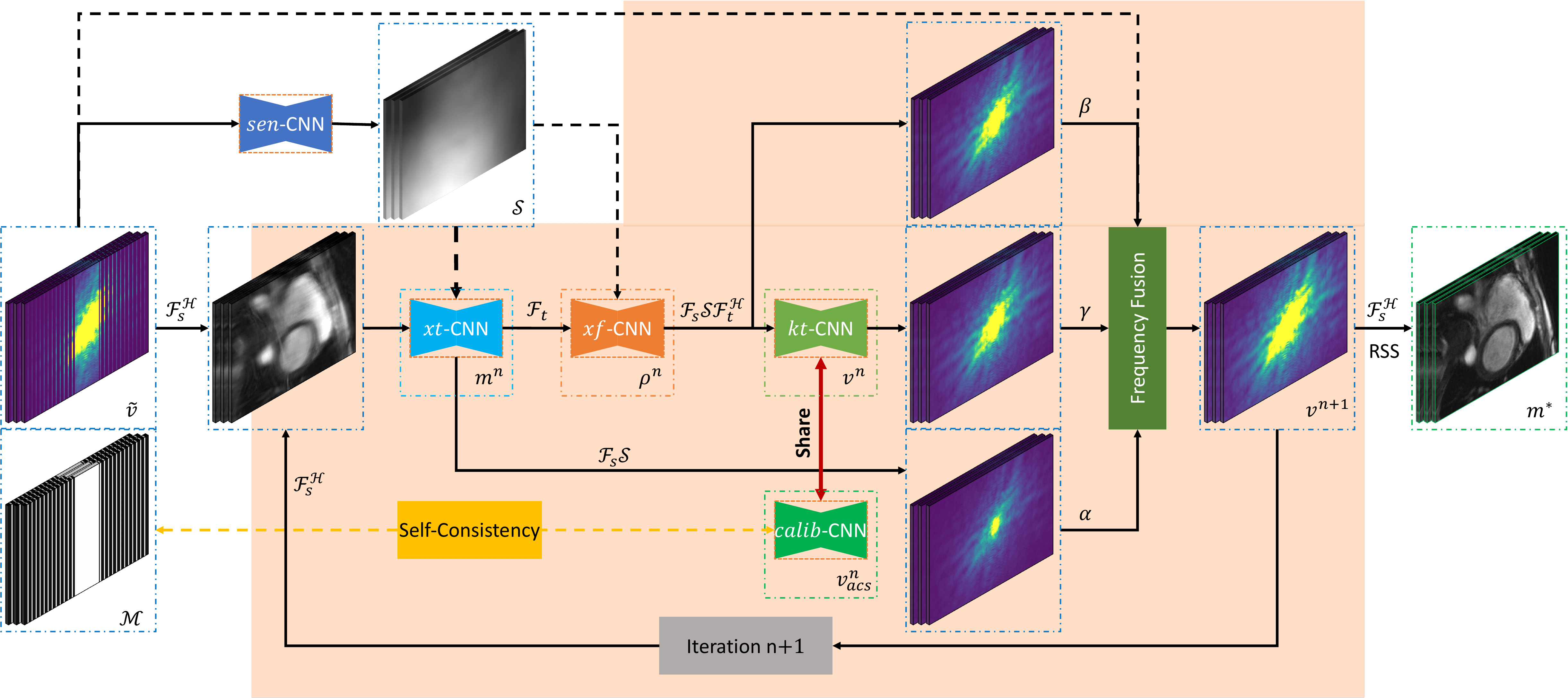}
	\caption{Illustration of the overall architecture of the proposed $k$-$t$ CLAIR.} \label{fig:ktclair}
\end{figure}
We propose $k$-$t$ CLAIR, as shown in Fig.~\ref{fig:ktclair}, to tackle the optimization problem with multiple priors in Eq.~\eqref{eq2} through an iterative approach. $k$-$t$ CLAIR exploits spatiotemporal correlations by iteratively learning complementary priors in the $x$-$t$, ($x$-$f$), and $k$-$t$ domains while enforcing self-consistency learning in the $k$-$t$ domain.
Specifically, the proposed approach consists of four steps of prior learning within each iteration: (1) an image enhancement prior learning step removes aliasing artifacts in the $x$-$t$ domain using a $xt$-CNN; (2) a dynamic temporal prior learning step captures dynamic details in the $x$-$f$ domain through a $xf$-CNN; (3) a k-space restoration prior learning step restores high-frequency information in the $k$-$t$ domain using a $kt$-CNN; and (4) a self-consistency prior learning step enforces calibration data consistency in the $k$-$t$ domain via a $calib$-CNN. Furthermore, we introduce a frequency fusion block to merge and coordinate the feature learning processes of all the priors. Notably, in our approach, we also incorporate the joint learning of coil sensitivity maps using a $sen$-CNN, inspired by the work~\cite{zbontar2018fastmri}. This joint learning of coil sensitivity maps complements the multi-prior learning framework of $k$-$t$ CLAIR and enhances the overall reconstruction process for dynamic parallel MRI.

\subsubsection{Image enhancement prior in the $x$-$t$ domain}
We begin by leveraging the spatiotemporal correlations present in the coil-combined image sequence in the $x$-$t$ domain during each iteration. This is achieved by learning a data-adaptive image prior from the training data using an $xt$-CNN. Our objective is to generate artifact-free images and restore the overall anatomical structure from highly degraded images, enabling us to provide complete k-space signals for subsequent dynamic feature extraction and high-frequency restoration. Specifically, within each iteration, the $xt$-CNN employs a step of the unrolled gradient descent (GD) algorithm to guarantee fidelity to the acquired data and enforce image constraints on the coil-combined images. At the $n$-th iteration, the $x$-$t$ domain reconstruction can be expressed as follows:
\begin{equation}
	\mathbf{m}^{n+1} =
	\mathbf{m}^{n} - \eta^{n} \big(
	xt\text{-CNN}^n(\mathbf{m}^n) + \lambda^n \mathcal{S}^H\mathcal{F}^H_\text{s}(\mathcal{M}\mathcal{F}_\text{s}\mathcal{S}\mathbf{m}^{n} - \tilde{\mathbf{v}})
	\big),
\end{equation}
where $\eta^{n}$ denotes the learning rate at the $n$-th iteration. Similar to the approach in~\cite{sriram2020end}, we utilize U-Net~\cite{ronneberger2015u} to learn a highly nonlinear image prior.

\subsubsection{Dynamic temporal prior in the $x$-$f$ domain}
To further capture dynamic features, we propose to learn a temporal prior in the $x$-$f$ domain. This approach takes advantage of the sparsity property exhibited by the signal in the temporal Fourier domain, arising from the periodic cardiac motion observed in dynamic imaging.
Our $xf$-CNN, similar to the $xt$-CNN, utilizes the GD algorithm for data fidelity and capturing temporal dynamic frequency. Compared to methods in~\cite{jung2009k,qin2021complementary}, our direct restoration of temporal dynamic frequencies proves more effective in restoring dynamic motion patterns within our framework.
At the $n$-th iteration, the $x$-$f$ domain reconstruction can be formulated as follows:
\begin{equation}
	\boldsymbol{\rho}^{n+1} =
	\boldsymbol{\rho}^{n} - \zeta^{n} \big(
	xf\text{-CNN}^n(\boldsymbol{\rho}^n) + \lambda^n \mathcal{F}_\text{t}\mathcal{S}^H\mathcal{F}^H_\text{s}(\mathcal{M}\mathcal{F}_\text{s}\mathcal{S}\mathcal{F}^H_\text{t}\boldsymbol{\rho}^{n} - \tilde{\mathbf{v}})
	\big),
\end{equation}
where $\zeta^{n}$ denotes the learning rate. In our approach, we leverage U-Net~\cite{ronneberger2015u} to learn a highly nonlinear dynamic temporal prior.

\subsubsection{Self-Consistency guided prior in the $k$-$t$ domain}
Previous approaches~\cite{sriram2020grappanet,ryu2021k,zhang2023camp} have shown that leveraging k-space correlations preserves high-frequency information and enhances image quality. We extend this idea by employing a $kt$-CNN model in the $k$-$t$ domain to learn spatio-temporal k-space priors from multi-coil data for high-frequency restoration. The $kt$-CNN consists of four convolutional layers that estimate missing signals based on neighboring data.
To ensure more accurate and consistent signal restoration in the $k$-$t$ domain, we incorporate calibration information using a $calib$-CNN. The $calib$-CNN shares the same architecture as the $kt$-CNN and learns scan-specific feature embeddings from auto-calibration signals (ACS). By sharing network parameters between the $calib$-CNN and $kt$-CNN models, we enable the learning of scan-specific features that improve the prediction of missing high-frequency k-space data. This integration of calibration information enhances reconstruction accuracy and fidelity in the $k$-$t$ domain. The update formula for the $n$-th iteration can be expressed as follows:
\begin{equation}
	\mathbf{v}^{n+1}_\text{acs} = calib\text{-CNN}^{n}(\mathbf{v}_\text{acs}), \quad
	\mathbf{v}^{n+1} = kt\text{-CNN}^{n}(\mathbf{v}^{n}),
\end{equation}
where $\mathbf{v}_\text{acs}$ refers to the ACS data.

\subsubsection{Frequency fusion block}
To leverage the benefits of different priors, a frequency fusion layer is introduced to coordinate the feature learning processes in the $x$-$t$, $x$-$f$, and $k$-$t$ domains. The formulation of this block is as follows:
\begin{equation}
	\mathbf{v}^{n+1} = \alpha\mathcal{MF}_\text{s}\mathcal{S}\mathbf{m}^{n} + \beta\mathcal{MF}_\text{s}\mathcal{S}\mathcal{F}_\text{t}^H\boldsymbol{\rho}^{n} + \gamma(1-\mathcal{M})\mathbf{v}^{n},
\end{equation}
where the coefficients $\alpha$, $\beta$, and $\gamma$ control the influence of each prior in the fusion process. By adjusting these coefficients, the network can balance the contributions from different priors based on their respective strengths and leverage their complementary information for more accurate and faithful dynamic MRI reconstruction. For simplicity during training, the values of $\alpha$, $\beta$, and $\gamma$ were initially set to 0.5, 0.5, and 1, respectively. However, further exploration can be conducted by treating these values as learnable parameters.

\subsection{Objective Function and Evaluation Metric}
For quantitative evaluation, we employ three commonly used reconstruction metrics: Structural Similarity Index (SSIM), Normalized Mean Squared Error (NMSE), and Peak Signal-to-Noise Ratio (PSNR). These metrics allow us to assess the quality and fidelity of the reconstructed data. During the training process, we utilize two loss functions: $L_1$ loss and SSIM loss. These loss functions are employed to minimize the difference between the reconstructed images and the ground truth, thereby ensuring accurate reconstruction. Additionally, we apply an $L_1$ loss to constrain the $calib$-CNN model throughout training. This constraint helps maintain consistent embedding of acquisition features in each iteration, thereby preserving the integrity of the acquired data during the reconstruction process. The total losses can be expressed as follows:
\begin{equation}
	\mathcal{L} = \lambda_1\mathcal{L}_{L_1}(m^\ast, m^G) + \lambda_2\mathcal{L}_{SSIM}(m^\ast,m^G) + \lambda_3\sum^{T}_{n=1}\mathcal{L}_{1}(v^{n}_\text{acs},v_\text{acs}),
\end{equation}
where $m^\ast$ and $m^G$ are the Root Sum of Squares (RSS) reconstruction and reference image, respectively. $v^{n}_\text{acs}$ represents the predicted ACS data from the $(n-1)$-th $calib$-CNN using the acquired ACS data $v_\text{acs}$ as input. $\lambda_1$, $\lambda_2$, and $\lambda_3$ are trade-off parameters set to 1 during training. 

\section{Experiment and Results}
\subsection{Data and Implementation Details}
We conducted experiments on two Cardiac MRI Reconstruction tasks: accelerated cine and T1/T2 mapping, as part of the CMRxRecon Challenge. The dataset included 300 healthy volunteers from a single center, with 120 training data, 60 validation data, and 120 test data. The training data consisted of fully sampled k-space raw data, while the validation and test data consisted of undersampled k-space data with acceleration factors of 4, 8, and 10, along with a sampling mask and ACS of 24 lines. The ground truth images for the validation set were withheld, and the test data was not accessible to the participants. For more detailed information, please refer to~\cite{wang2021recommendation,wang2023cmrxrecon} and the challenge website\footnote{\url{https://cmrxrecon.github.io}}.

\subsection{Implementation Details}
We divided the training data into two subsets: 80\% for model training and the remaining 20\% for model validation. All models were optimized using the Adam optimizer with parameters $\beta_1 = 0.9$ and $\beta_2 = 0.999$, an initial learning rate of $3e^{-4}$, and a batch size of 1. The number of iterations ($T$) was set to 12 for all rolling-based models. The network architecture remained consistent for both the $xt$-CNN and $xf$-CNN models. For the mapping task, the initial feature dimension was set to 64, and for the cine task, it was set to 32. These dimensions were then doubled after each pooling stage to enhance the network's ability to capture critical features. Regarding the $kt$-CNN and $calib$-CNN models, the feature dimension was specifically set to match the number of coils in each respective task. Networks were trained for 30 epochs for the cine reconstruction task and 50 epochs for the T1/T2 mapping reconstruction task. After 20 and 40 epochs, the learning rates were reduced by a factor of 10. Models were trained using PyTorch Lightning on 2 NVIDIA RTX A6000 GPUs for T1/T2 mapping and 4 Tesla A100 GPUs for the Cine task. Detailed model information for both tasks can be found in the Supplementary Materials.

\subsection{Experiments and Results}
\subsubsection{Reconstruction of Accelerated T1/T2 Acquisitions for T1/T2 mapping}
\begin{figure}[!t]
	\centering
	\includegraphics[width=0.9\textwidth]{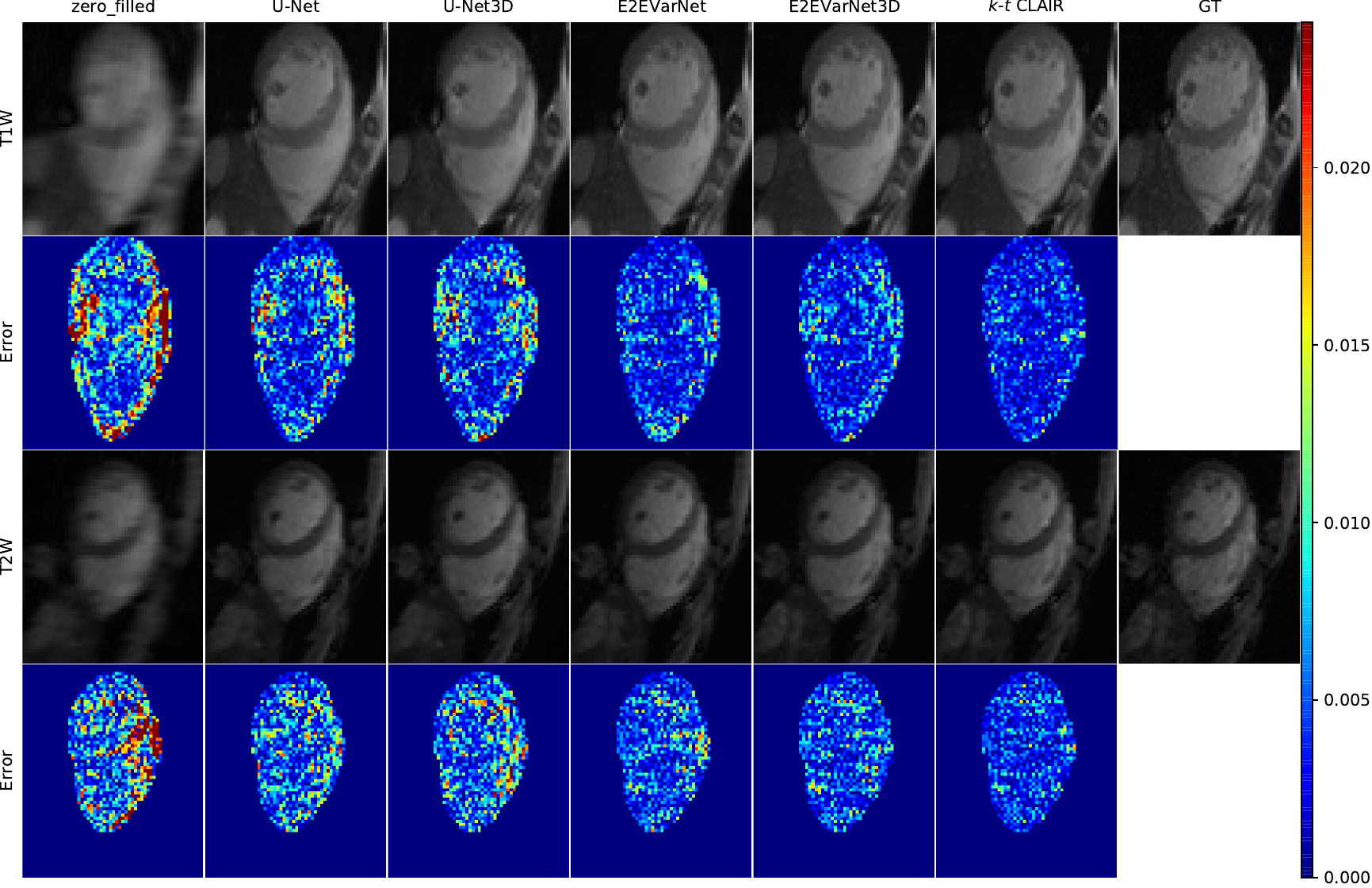}
	\caption{Reconstruction of 10$\times$ accelerated T1W/T2W images and masked error maps.} \label{fig:mapping}
\end{figure}
\begin{figure}[!t]
	\centering
	\includegraphics[width=0.9\textwidth]{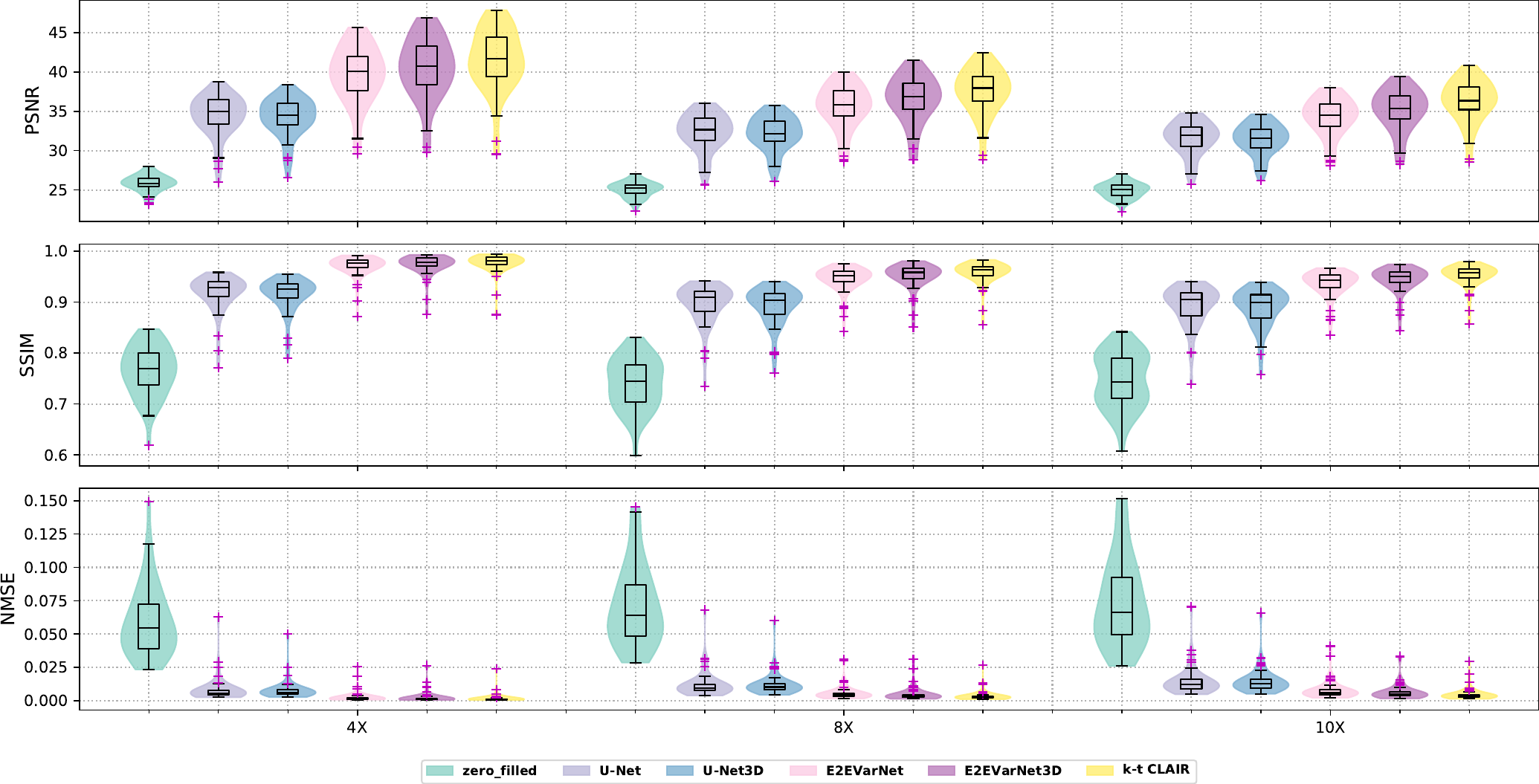}
	\caption{Performance for T1W/T2W reconstruction on the unseen training data.} \label{fig:mapping-violinbox}
\end{figure}
We conduct a comparative analysis of different methods for reconstruction of accelerated acquisitions of T1- and T2-weighted images in T1 and T2 mapping, respectively. The evaluated methods include Zero-filled, FastMRI U-Net~\cite{zbontar2018fastmri}, and E2EVarNet~\cite{sriram2020end}. Additionally, we extend the works in~\cite{zbontar2018fastmri,sriram2020end} by incorporating 3D Convolution to capture temporal correlations. Notably, E2EVarNet~\cite{sriram2020end} utilizes a deep learning unrolling network structure for iterative reconstruction in the image domain, while our proposed method is based on an unrolling framework that incorporates multiple prior knowledge in each reconstruction iteration. We demonstrate the effectiveness of our proposed method by presenting reconstruction results for highly accelerated T1 and T2-weighted images, as shown in Fig.~\ref{fig:mapping}. Our method yields reconstructions with enhanced fidelity, accurately capturing rich and intricate structures compared to alternative methods. The reconstruction error maps consistently highlight the superior performance of our approach in restoring highly undersampled images in the ventricle and ventricular myocardium regions. Detailed comparison results on the 20\% hold-out training data are presented in Table~\ref{tab:trainingset}, demonstrating the superiority of our method over other approaches for both T1 and T2-weighted images across all acceleration rates and evaluation metrics. Fig.~\ref{fig:mapping-violinbox} further illustrates the effectiveness of the proposed method. Additionally, $k$-$t$ CLAIR demonstrates strong generalization capabilities and achieves promising reconstructions on validation data. This is supported by the detailed quantitative scores obtained from the Challenge Leaderboard\footnote{\url{https://www.synapse.org/\#!Synapse:syn51471091/wiki/622548}} for cardiac T1 and T2-weighted reconstruction on the validation set, as shown in Table~\ref{tab:validationset}.

\begin{table}[!t]
	\centering
	\caption{Quantitative evaluation for cardiac CINE (LAX and SAX) and T1W/T2W images reconstruction on the unseen training data. Note that the results were reported after cropping the middle 1/6 of the original image.}\label{tab:trainingset}
	\resizebox{\textwidth}{!}{
			\begin{tabular}{l | c | c|c|c|c | c|c|c|c | c|c|c|c}
					\toprule
					\multirow{2}{*}{Methods} & \multirow{2}{*}{Acc.} & \multicolumn{4}{c|}{SSIM$\uparrow$} & \multicolumn{4}{c|}{NMSE$\downarrow$} & \multicolumn{4}{c}{PSNR$\uparrow$}\\
					\cmidrule(r){3-6} \cmidrule(r){7-10} \cmidrule(r){11-14}
					& & Lax & Sax & T1W & T2W & Lax & Sax & T1W & T2W & Lax & Sax & T1W & T2W \\
					\midrule
					\multirow{4}{*}{Zero-filled}
					& 4$\times$ & 0.6975 & 0.7601 & 0.7301 & 0.7991 & 0.0677 & 0.0599 & 0.0769 & 0.0417 & 26.06 & 27.22 & 25.92 & 25.79 \\
					& 8$\times$ & 0.6902 & 0.7328 & 0.7022 & 0.7771 & 0.0756 & 0.0738 & 0.0910 & 0.0507 & 25.82 & 26.40 & 25.17 & 24.95 \\
					&10$\times$ & 0.6864 & 0.7284 & 0.7023 & 0.7901 & 0.0788 & 0.0780 & 0.0963 & 0.0509 & 25.65 & 26.16 & 24.93 & 24.96 \\
					\cmidrule(r){3-14}
					& Avg. & \multicolumn{2}{c|}{0.7188} & \multicolumn{2}{c|}{0.7501} & \multicolumn{2}{c|}{0.0721} & \multicolumn{2}{c|}{0.0679} & \multicolumn{2}{c|}{26.26} & \multicolumn{2}{c}{25.29} \\
					\midrule
					\multirow{4}{*}{U-Net}
					& 4$\times$ & 0.8503 & 0.8983 & 0.9109 & 0.9247 & 0.0197 & 0.0123 & 0.0106 & 0.0064 & 31.24 & 34.17 & 35.05 & 33.74 \\
					& 8$\times$ & 0.8262 & 0.8701 & 0.8876 & 0.9053 & 0.0299 & 0.0202 & 0.0151 & 0.0098 & 29.62 & 31.97 & 33.13 & 31.66 \\
					&10$\times$ & 0.8152 & 0.8635 & 0.8809 & 0.9055 & 0.0338 & 0.0229 & 0.0186 & 0.0111 & 29.04 & 31.39 & 32.08 & 31.11 \\
					\cmidrule(r){3-14}
					& Avg. & \multicolumn{2}{c|}{0.8566} & \multicolumn{2}{c|}{0.9025} & \multicolumn{2}{c|}{0.0226} & \multicolumn{2}{c|}{0.0119} & \multicolumn{2}{c|}{31.39} & \multicolumn{2}{c}{32.79} \\
					\midrule
					\multirow{4}{*}{U-Net3D}
					& 4$\times$ & 0.8604 & 0.9076 & 0.9078 & 0.9216 & 0.0175 & 0.0106 & 0.0097 & 0.0065 & 31.64 & 34.72 & 34.92 & 33.49 \\
					& 8$\times$ & 0.8365 & 0.8798 & 0.8845 & 0.9017 & 0.0266 & 0.0179 & 0.0147 & 0.0100 & 29.99 & 32.44 & 32.94 & 31.45 \\
					&10$\times$ & 0.8262 & 0.8728 & 0.8770 & 0.9024 & 0.0305 & 0.0207 & 0.0183 & 0.0112 & 29.40 & 31.80 & 31.95 & 30.99 \\
					\cmidrule(r){3-14}
					& Avg. & \multicolumn{2}{c|}{0.8665} & \multicolumn{2}{c|}{0.8992} & \multicolumn{2}{c|}{0.0202} & \multicolumn{2}{c|}{0.0117} & \multicolumn{2}{c|}{31.82} & \multicolumn{2}{c}{32.63} \\
					\midrule
					\multirow{4}{*}{E2EVarNet}
					& 4$\times$ & 0.9569 & 0.9714 & 0.9738 & 0.9661 & 0.0039 & 0.0027 & 0.0035 & 0.0024 & 38.16 & 41.14 & 41.26 & 37.94 \\
					& 8$\times$ & 0.9016 & 0.9324 & 0.9466 & 0.9423 & 0.0130 & 0.0083 & 0.0071 & 0.0052 & 32.96 & 35.87 & 36.73 & 34.43 \\
					&10$\times$ & 0.8849 & 0.9183 & 0.9342 & 0.9378 & 0.0166 & 0.0109 & 0.0103 & 0.0064 & 31.92 & 34.67 & 34.97 & 33.52 \\
					\cmidrule(r){3-14}
					& Avg. & \multicolumn{2}{c|}{0.9291} & \multicolumn{2}{c|}{0.9501} & \multicolumn{2}{c|}{0.0090} & \multicolumn{2}{c|}{0.0058} & \multicolumn{2}{c|}{35.95} & \multicolumn{2}{c}{36.48} \\
					\midrule
					\multirow{4}{*}{E2EVarNet3D}
					& 4$\times$ & 0.9640 & 0.9759 & 0.9771 & 0.9692 & 0.0031 & 0.0022 & 0.0030 & 0.0020 & 39.11 & 42.01 & 42.34 & 38.61 \\
					& 8$\times$ & 0.9156 & 0.9429 & 0.9538 & 0.9485 & 0.0106 & 0.0067 & 0.0059 & 0.0042 & 33.88 & 36.83 & 37.78 & 35.26 \\
					&10$\times$ & 0.9003 & 0.9297 & 0.9434 & 0.9446 & 0.0138 & 0.0090 & 0.0085 & 0.0052 & 32.77 & 35.54 & 36.00 & 34.37 \\
					\cmidrule(r){3-14}
					& Avg. & \multicolumn{2}{c|}{0.9394} & \multicolumn{2}{c|}{0.9561} & \multicolumn{2}{c|}{0.0074} & \multicolumn{2}{c|}{0.0048} & \multicolumn{2}{c|}{36.86} & \multicolumn{2}{c}{37.39} \\
					\midrule
					\multirow{4}{*}{$k$-$t$ CLAIR}
					& 4$\times$ & 0.9673 & 0.9777 & 0.9806 & 0.9728 & 0.0027 & 0.0020 & 0.0021 & 0.0015 & 39.74 & 42.49 & 43.57 & 39.48 \\
					& 8$\times$ & 0.9238 & 0.9493 & 0.9609 & 0.9544 & 0.0090 & 0.0056 & 0.0043 & 0.0032 & 34.57 & 37.64 & 38.90 & 36.14 \\
					&10$\times$ & 0.9099 & 0.9372 & 0.9531 & 0.9507 & 0.0118 & 0.0075 & 0.0060 & 0.0040 & 33.47 & 36.30 & 37.31 & 35.19 \\
					\cmidrule(r){3-14}
					& Avg. & \multicolumn{2}{c|}{\textbf{0.9454}} & \multicolumn{2}{c|}{\textbf{0.9621}} & \multicolumn{2}{c|}{\textbf{0.0063}} & \multicolumn{2}{c|}{\textbf{0.0035}} & \multicolumn{2}{c|}{\textbf{37.54}}& \multicolumn{2}{c}{\textbf{38.43}} \\
					\bottomrule
				\end{tabular}
		}
\end{table}

\begin{table}[!t]
	\centering
	\caption{Quantitative evaluation of the proposed $k$-$t$ CLAIR for cardiac CINE (LAX and SAX) and T1W/T2W images reconstruction on the Validation Set.}\label{tab:validationset}
	\resizebox{\textwidth}{!}{
			\begin{tabular}{l | c|c|c|c | c|c|c|c | c|c|c|c}
					\toprule
					\multirow{2}{*}{Acc.}
					& \multicolumn{4}{c|}{SSIM$\uparrow$} & \multicolumn{4}{c|}{NMSE$\downarrow$} & \multicolumn{4}{c}{PSNR$\uparrow$}\\
					\cmidrule(r){2-5} \cmidrule(r){6-9} \cmidrule(r){10-13}
					& Lax & Sax & T1W & T2W & Lax & Sax & T1W & T2W & Lax & Sax & T1W & T2W \\
					\midrule
					\multirow{1}{*}{4$\times$}
					& 0.963 & 0.976 & 0.983 & 0.970 & 0.003 & 0.003 & 0.003 & 0.003 & 38.838 & 41.656 & 43.801 & 38.849 \\
					\midrule
					\multirow{1}{*}{8$\times$}
					& 0.914 & 0.945 & 0.961 & 0.950 & 0.011 & 0.007 & 0.006 & 0.006 & 33.508 & 36.853 & 38.961 & 35.621 \\
					\midrule
					\multirow{1}{*}{10$\times$}
					& 0.901 & 0.932 & 0.954 & 0.947 & 0.014 & 0.009 & 0.008 & 0.007 & 32.49 & 35.602 & 37.225 & 34.721 \\
					\midrule
					\multirow{1}{*}{Avg.}
					& \multicolumn{2}{c|}{0.939} & \multicolumn{2}{c|}{0.961} & \multicolumn{2}{c|}{0.008} & \multicolumn{2}{c|}{0.005} & \multicolumn{2}{c|}{36.491} & \multicolumn{2}{c}{38.196} \\
					\bottomrule
				\end{tabular}
		}
\end{table}
 
\subsubsection{Accelerated Cine Images Reconstruction}
\begin{figure}[!t]
	\centering
	\includegraphics[width=0.9\textwidth]{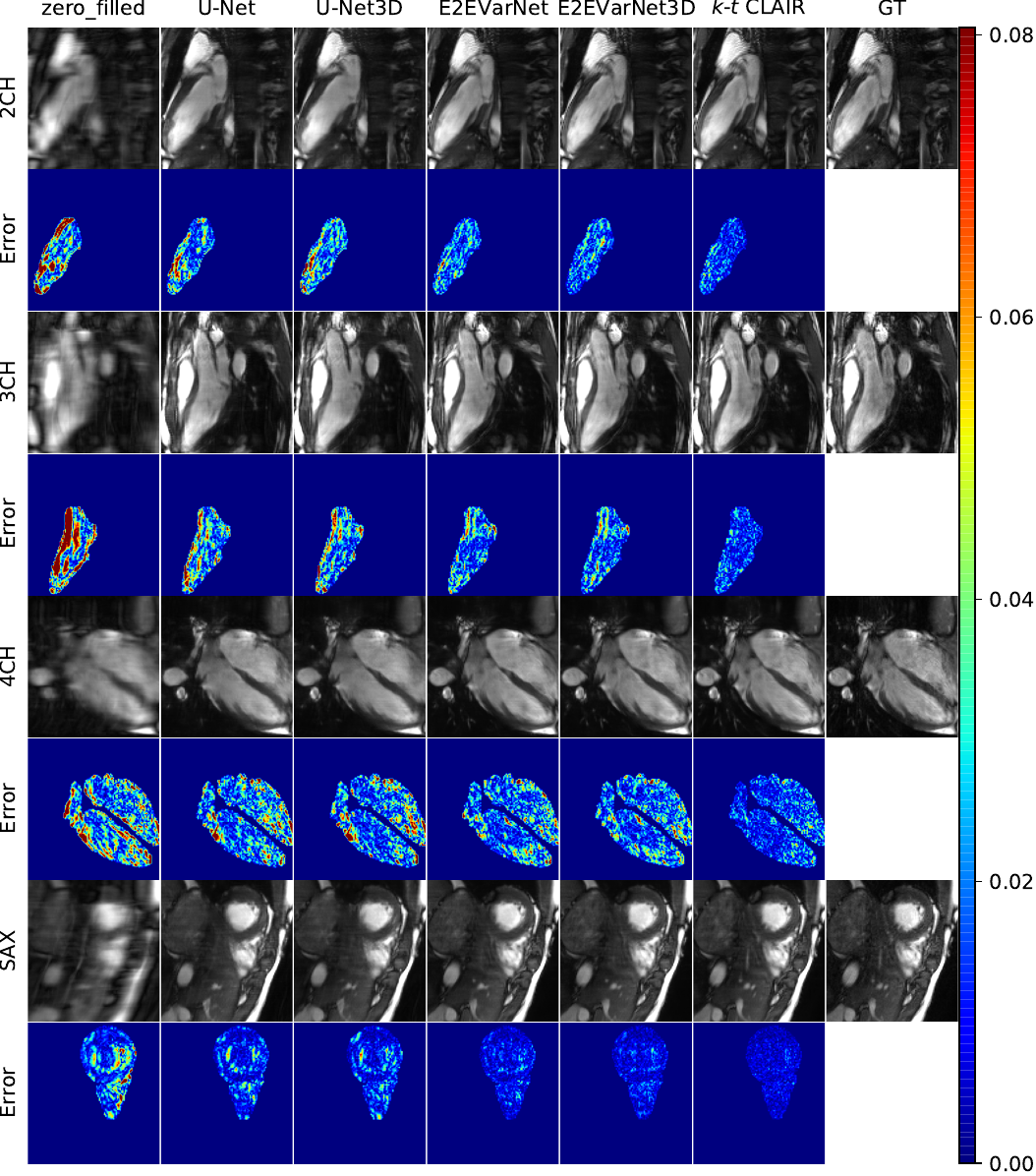}
	\caption{Reconstruction of 10$\times$ accelerated CINE and masked error maps.} \label{fig:cine}
\end{figure}
\begin{figure}[!t]
	\centering
	\includegraphics[width=0.9\textwidth]{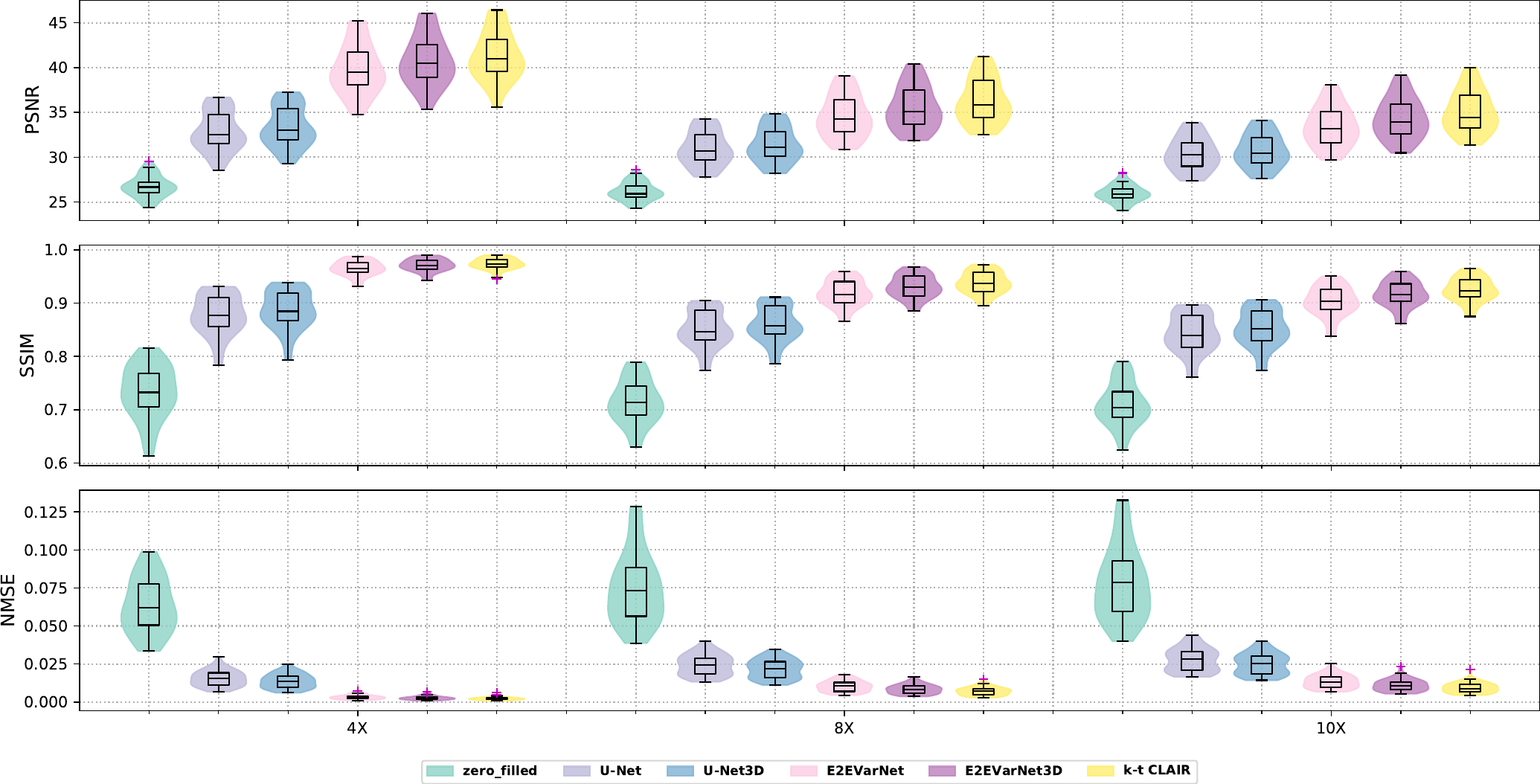}
	\caption{Performance for CINE reconstruction on the unseen training data.} \label{fig:cine-violinbox}
\end{figure}
For the cine task evaluation, we compare our method with Zero-filled, FastMRI U-Net~\cite{zbontar2018fastmri}, and E2EVarNet~\cite{sriram2020end}. Additionally, we compare our method with U-Net3D and E2EVarNet3D, similar to the mapping task. The reconstruction results for 10$\times$ acceleration are shown in Fig.~\ref{fig:cine}, illustrating the faithful reconstructions produced by our method across all views, in comparison to other methods. The reconstruction error maps within the myocardium and chambers consistently demonstrate the superior accuracy of our method in rapid CMR reconstruction. Quantitative results on the 20\% hold-out training data are presented in Table~\ref{tab:trainingset}, indicating the outstanding reconstruction quality and improvement achieved by our proposed method across all evaluation metrics and views. The effectiveness of the proposed method is consistently illustrated by the violin plot, as shown in Fig.~\ref{fig:cine-violinbox}. The performance on the Sax view generally surpasses that of the Lax views at the same acceleration rate, possibly due to the larger amount of available data for the Sax view. Furthermore, our method also performs well in generating reconstructions for cine data in the validation set, as detailed in Table~\ref{tab:validationset}.

\section{Conclusions}
In this paper, we introduced $k$-$t$ CLAIR, a self-consistency guided multi-prior learning framework for faithful dynamic MRI reconstruction. By leveraging multiple priors and incorporating calibration information, $k$-$t$ CLAIR improved the quality and accuracy of dynamic MRI reconstructions. Experimental results demonstrated its effectiveness in achieving high-quality reconstructions for cardiac cine and T1W/T2W images in T1/T2 mapping. In future work, we plan to extend the application of this method to other anatomical structures in dynamic MRI and validate its application in clinical routines.

\subsubsection{Acknowledgements.}
This work was supported by a grant from Innovation and Technology Commission of the Hong Kong SAR (MRP/046/20X); and by a Faculty Innovation Award from the Faculty of Medicine of The Chinese University of Hong Kong.

%
%
 \bibliographystyle{splncs04}
 \bibliography{ref}

\begin{thebibliography}{10}
\providecommand{\url}[1]{\texttt{#1}}
\providecommand{\urlprefix}{URL }
\providecommand{\doi}[1]{https://doi.org/#1}

\bibitem{aggarwal2018modl}
Aggarwal, H.K., Mani, M.P., Jacob, M.: Modl: Model-based deep learning
  architecture for inverse problems. IEEE transactions on medical imaging
  \textbf{38}(2),  394--405 (2018)

\bibitem{akccakaya2019scan}
Ak{\c{c}}akaya, M., Moeller, S., Weing{\"a}rtner, S., U{\u{g}}urbil, K.:
  Scan-specific robust artificial-neural-networks for k-space interpolation
  (raki) reconstruction: Database-free deep learning for fast imaging. Magnetic
  resonance in medicine  \textbf{81}(1),  439--453 (2019)

\bibitem{eo2018kiki}
Eo, T., Jun, Y., Kim, T., Jang, J., Lee, H.J., Hwang, D.: Kiki-net:
  cross-domain convolutional neural networks for reconstructing undersampled
  magnetic resonance images. Magnetic resonance in medicine  \textbf{80}(5),
  2188--2201 (2018)

\bibitem{fabian2022humus}
Fabian, Z., Tinaz, B., Soltanolkotabi, M.: Humus-net: Hybrid unrolled
  multi-scale network architecture for accelerated mri reconstruction. Advances
  in Neural Information Processing Systems  \textbf{35},  25306--25319 (2022)

\bibitem{griswold2002generalized}
Griswold, M.A., Jakob, P.M., Heidemann, R.M., Nittka, M., Jellus, V., Wang, J.,
  Kiefer, B., Haase, A.: Generalized autocalibrating partially parallel
  acquisitions (grappa). Magnetic Resonance in Medicine: An Official Journal of
  the International Society for Magnetic Resonance in Medicine  \textbf{47}(6),
   1202--1210 (2002)

\bibitem{hammernik2018learning}
Hammernik, K., Klatzer, T., Kobler, E., Recht, M.P., Sodickson, D.K., Pock, T.,
  Knoll, F.: Learning a variational network for reconstruction of accelerated
  mri data. Magnetic resonance in medicine  \textbf{79}(6),  3055--3071 (2018)

\bibitem{han2020k}
Han, Y., Sunwoo, L., Ye, J.C.: k-space deep learning for accelerated mri. IEEE
  transactions on medical imaging  \textbf{39}(2),  377--386 (2020)

\bibitem{jung2009k}
Jung, H., Sung, K., Nayak, K.S., Kim, E.Y., Ye, J.C.: k-t focuss: a general
  compressed sensing framework for high resolution dynamic mri. Magnetic
  Resonance in Medicine: An Official Journal of the International Society for
  Magnetic Resonance in Medicine  \textbf{61}(1),  103--116 (2009)

\bibitem{kustner2020cinenet}
K{\"u}stner, T., Fuin, N., Hammernik, K., Bustin, A., Qi, H., Hajhosseiny, R.,
  Masci, P.G., Neji, R., Rueckert, D., Botnar, R.M., et~al.: Cinenet: deep
  learning-based 3d cardiac cine mri reconstruction with multi-coil
  complex-valued 4d spatio-temporal convolutions. Scientific reports
  \textbf{10}(1),  13710 (2020)

\bibitem{lingala2011accelerated}
Lingala, S.G., Hu, Y., DiBella, E., Jacob, M.: Accelerated dynamic mri
  exploiting sparsity and low-rank structure: kt slr. IEEE transactions on
  medical imaging  \textbf{30}(5),  1042--1054 (2011)

\bibitem{lustig2007sparse}
Lustig, M., Donoho, D., Pauly, J.M.: Sparse mri: The application of compressed
  sensing for rapid mr imaging. Magnetic Resonance in Medicine: An Official
  Journal of the International Society for Magnetic Resonance in Medicine
  \textbf{58}(6),  1182--1195 (2007)

\bibitem{lustig2010spirit}
Lustig, M., Pauly, J.M.: Spirit: iterative self-consistent parallel imaging
  reconstruction from arbitrary k-space. Magnetic resonance in medicine
  \textbf{64}(2),  457--471 (2010)

\bibitem{otazo2015low}
Otazo, R., Candes, E., Sodickson, D.K.: Low-rank plus sparse matrix
  decomposition for accelerated dynamic mri with separation of background and
  dynamic components. Magnetic resonance in medicine  \textbf{73}(3),
  1125--1136 (2015)

\bibitem{pruessmann1999sense}
Pruessmann, K.P., Weiger, M., Scheidegger, M.B., Boesiger, P.: Sense:
  sensitivity encoding for fast mri. Magnetic Resonance in Medicine: An
  Official Journal of the International Society for Magnetic Resonance in
  Medicine  \textbf{42}(5),  952--962 (1999)

\bibitem{qin2021complementary}
Qin, C., Duan, J., Hammernik, K., Schlemper, J., K{\"u}stner, T., Botnar, R.,
  Prieto, C., Price, A.N., Hajnal, J.V., Rueckert, D.: Complementary
  time-frequency domain networks for dynamic parallel mr image reconstruction.
  Magnetic Resonance in Medicine  \textbf{86}(6),  3274--3291 (2021)

\bibitem{qin2018convolutional}
Qin, C., Schlemper, J., Caballero, J., Price, A.N., Hajnal, J.V., Rueckert, D.:
  Convolutional recurrent neural networks for dynamic mr image reconstruction.
  IEEE transactions on medical imaging  \textbf{38}(1),  280--290 (2018)

\bibitem{qin2019k}
Qin, C., Schlemper, J., Duan, J., Seegoolam, G., Price, A., Hajnal, J.,
  Rueckert, D.: k-t next: dynamic mr image reconstruction exploiting
  spatio-temporal correlations. In: Medical Image Computing and Computer
  Assisted Intervention--MICCAI 2019: 22nd International Conference, Shenzhen,
  China, October 13--17, 2019, Proceedings, Part II 22. pp. 505--513. Springer
  (2019)

\bibitem{ran2020md}
Ran, M., Xia, W., Huang, Y., Lu, Z., Bao, P., Liu, Y., Sun, H., Zhou, J.,
  Zhang, Y.: Md-recon-net: A parallel dual-domain convolutional neural network
  for compressed sensing mri. IEEE Transactions on Radiation and Plasma Medical
  Sciences  \textbf{5}(1),  120--135 (2020)

\bibitem{ronneberger2015u}
Ronneberger, O., Fischer, P., Brox, T.: U-net: Convolutional networks for
  biomedical image segmentation. In: Medical Image Computing and
  Computer-Assisted Intervention--MICCAI 2015: 18th International Conference,
  Munich, Germany, October 5-9, 2015, Proceedings, Part III 18. pp. 234--241.
  Springer (2015)

\bibitem{ryu2021k}
Ryu, K., Alkan, C., Choi, C., Jang, I., Vasanawala, S.: K-space refinement in
  deep learning mr reconstruction via regularizing scan specific spirit-based
  self consistency. In: Proceedings of the IEEE/CVF International Conference on
  Computer Vision. pp. 4008--4017 (2021)

\bibitem{schlemper2017deep}
Schlemper, J., Caballero, J., Hajnal, J.V., Price, A.N., Rueckert, D.: A deep
  cascade of convolutional neural networks for dynamic mr image reconstruction.
  IEEE transactions on Medical Imaging  \textbf{37}(2),  491--503 (2017)

\bibitem{sriram2020end}
Sriram, A., Zbontar, J., Murrell, T., Defazio, A., Zitnick, C.L., Yakubova, N.,
  Knoll, F., Johnson, P.: End-to-end variational networks for accelerated mri
  reconstruction. In: Medical Image Computing and Computer Assisted
  Intervention--MICCAI 2020: 23rd International Conference, Lima, Peru, October
  4--8, 2020, Proceedings, Part II 23. pp. 64--73. Springer (2020)

\bibitem{sriram2020grappanet}
Sriram, A., Zbontar, J., Murrell, T., Zitnick, C.L., Defazio, A., Sodickson,
  D.K.: Grappanet: Combining parallel imaging with deep learning for multi-coil
  mri reconstruction. In: Proceedings of the IEEE/CVF Conference on Computer
  Vision and Pattern Recognition. pp. 14315--14322 (2020)

\bibitem{wang2021recommendation}
Wang, C., Li, Y., Lv, J., Jin, J., Hu, X., Kuang, X., Chen, W., Wang, H.:
  Recommendation for cardiac magnetic resonance imaging-based phenotypic study:
  imaging part. Phenomics  \textbf{1},  151--170 (2021)

\bibitem{wang2023cmrxrecon}
Wang, C., Lyu, J., Wang, S., Qin, C., Guo, K., Zhang, X., Yu, X., Li, Y., Wang,
  F., Jin, J., et~al.: Cmrxrecon: An open cardiac mri dataset for the
  competition of accelerated image reconstruction. arXiv preprint
  arXiv:2309.10836  (2023)

\bibitem{wang2022dimension}
Wang, S., Ke, Z., Cheng, H., Jia, S., Ying, L., Zheng, H., Liang, D.:
  Dimension: dynamic mr imaging with both k-space and spatial prior knowledge
  obtained via multi-supervised network training. NMR in Biomedicine
  \textbf{35}(4),  e4131 (2022)

\bibitem{yang2017dagan}
Yang, G., Yu, S., Dong, H., Slabaugh, G., Dragotti, P.L., Ye, X., Liu, F.,
  Arridge, S., Keegan, J., Guo, Y., et~al.: Dagan: Deep de-aliasing generative
  adversarial networks for fast compressed sensing mri reconstruction. IEEE
  transactions on medical imaging  \textbf{37}(6),  1310--1321 (2017)

\bibitem{zbontar2018fastmri}
Zbontar, J., Knoll, F., Sriram, A., Murrell, T., Huang, Z., Muckley, M.J.,
  Defazio, A., Stern, R., Johnson, P., Bruno, M., et~al.: fastmri: An open
  dataset and benchmarks for accelerated mri. arXiv preprint arXiv:1811.08839
  (2018)

\bibitem{zhang2023camp}
Zhang, L., Li, X., Chen, W.: Camp-net: Consistency-aware multi-prior network
  for accelerated mri reconstruction. arXiv preprint arXiv:2306.11238  (2023)

\end{thebibliography}

\end{document}